\documentclass{ws-procs9x6}

\begin{document}

\title{FISSION OF ACTINIDES AND SUPERHEAVY NUCLEI:\\
COVARIANT DENSITY FUNCTIONAL PERSPECTIVE}

\author{A. V. AFANASJEV$^*$}

\address{Department Physics and Astronomy, Mississippi State University,\\
Mississippi State, MS 39762, USA\\
$^*$E-mail: afansjev@erc.msstate.edu}

%\author{A. N. AUTHOR}
%
%\address{Group, Laboratory, Street,\\
%City, State ZIP/Zone, Country\\
%E-mail: an\_author@laboratory.com}

\begin{abstract}
 The current status of the application of covariant density functional
theory to the description of fission barriers in actinides and superheavy
nuclei is reviewed. The achievements and open problems are discussed.
\end{abstract}


\bodymatter

%%%%%%%%%%%%%%%%%%%%%%%%%
\section{Introduction}
%%%%%%%%%%%%%%%%%%%%%%%%%

  A study of the fission barrier heights $B_{f}$ of nuclei is 
motivated by the importance of this quantity for several physical 
phenomena. For example, the $r-$process of stellar nucleosynthesis 
depends (among other quantities such as masses and $\beta$-decay 
rates) on the fission barriers of very neutron-rich nuclei 
\cite{AT.99,MPR.01}. The population and survival
of hyperdeformed states at high spin also depends on the fission
barriers\cite{DPS.04,AA.08}. In addition, the physics of fission barriers is 
intimately connected with on-going search for new superheavy elements 
(SHE).
% which is motivated by the
%attempts to provide the answers for two open questions in nuclear structure,
%namely, the limits of the existence of atomic nuclei at large
%values of proton number and the location of the island of stability of superheavy
%nuclei and the next magic numbers (if any) beyond $Z=82$ and $N=126$. 
The probability for the formation of a SHE in a heavy-ion-fusion reaction 
is directly connected to the height of its fission barrier\cite{IOZ.02};
%which is a decisive quantity in the
%competition between neutron evaporation and fission of a compound nucleus
%in the process of its cooling. 
the large sensitivity of the cross section $\sigma$ for the synthesis of 
the fissioning nuclei on the barrier height $B_{f}$ also stresses 
a need for accurate calculations of this value. The survival of the
actinides and SHE against spontaneous fission depends on the fission 
barrier which is a measure of the stability of a nucleus reflected in 
the spontaneous fission lifetimes of these nuclei\cite{SP.07}.

  The recent progress in the description of fission barriers within
covariant density functional theory (CDFT\cite{VALR.05}) is briefly 
reviewed in the current manuscript. The comparison with the results
of other model calculations is presented. Open problems in
the study of fission barriers are also discussed.  

%%%%%%%%%%%%%%%%%%%%%%%%%%%%%%%%%%%%%%%%%%%%%%%%%%%%%%%%%%%%%%%%%%%%%%%%%%%%%
\begin{figure}[ht]
\begin{center}
\includegraphics[width=9.0cm,angle=0]{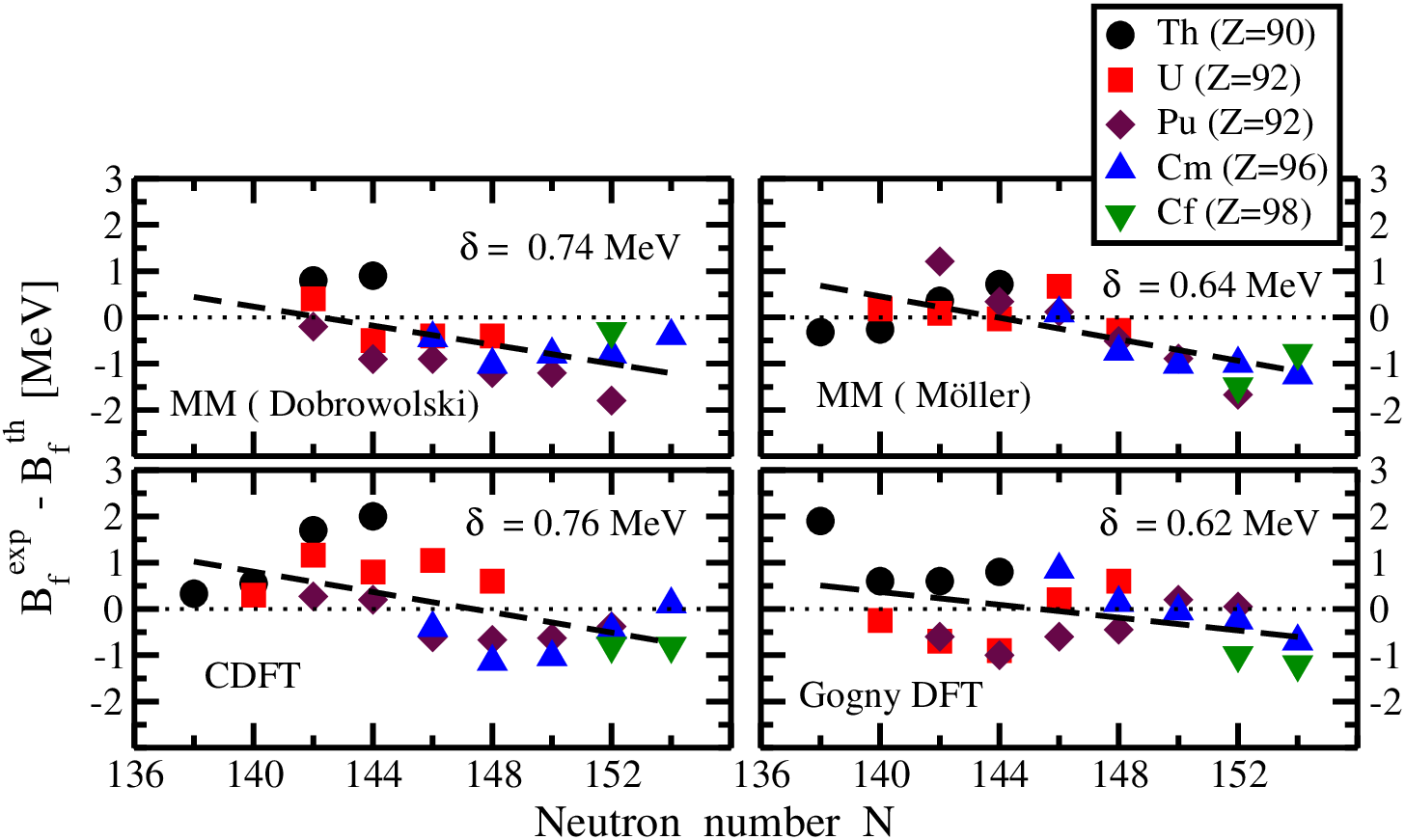}
\end{center}
\vspace{-0.1cm}
\caption{The difference between experimental and calculated heights 
of inner fission barriers as a function of neutron number $N$. The
results of the calculations are compared to estimated fission barrier
heights given in the RIPL-3 database \protect\cite{RIPL-3}, which is used for this 
purpose in the absolute majority of theoretical studies on fission barriers 
in actinides. The results of the calculations within microscopic+macroscopic
method ('MM(Dobrowolski)' \protect\cite{DPB.07} and 'MM(M{\"o}ller)' \protect\cite{MSI.09}), 
covariant density functional theory ('CDFT' \protect\cite{AAR.10}) and density 
functional theory based on the finite range Gogny force ('Gogny DFT' 
\protect\cite{DGGL.06}) are shown. Thick dashed lines are used to show the average 
trend of the deviations between theory and experiment as a function of 
neutron number. The average deviation per barrier $\delta$ [in MeV] is defined as 
$\delta = \sum_{i=1}^N |B_f^i(th)-B_f^i(exp)|/N$, where $N$ is the number 
of the barriers with known experimental heights, and $B_f^i(th)$ 
($B_f^i(exp)$) are calculated (experimental) heights of the barriers.
Long-dashed lines represent the trend of the deviations between theory and
experiment as a function of neutron number. They are obtained via linear
regression based on a least square fit. From Ref. \protect\cite{AAR.12}.}
\label{N-dep}
\end{figure}
%%%%%%%%%%%%%%%%%%%%%%%%%%%%%%%%%%%%%%%%%%%%%%%%%%%%%%%%%%%%%%%%%%%%%%%%%%%%%%%%
%%%%%%%%%%%%%%%%%%%%%%%%%%%%%%%%%%%%%%%%%%%%%%%%%%%%%%%%%%%%%%%%
\section{Current status of the investigation of fission barriers
in actinides}
%%%%%%%%%%%%%%%%%%%%%%%%%%%%%%%%%%%%%%%%%%%%%%%%%%%%%%%%%%%%%%%%

  The progress in the development of computer codes and the 
availability of powerful computers has allowed to study in a 
systematic way the effects of triaxiality on the fission barriers 
leading to their realistic description. Within the CDFT framework, 
the inner fission barriers with triaxiality included have been 
studied for the first time in Ref.\ \cite{AAR.10} using triaxial
RMF+BCS approach and the NL3* parametrization \cite{NL3*}. 
Two years later, 
similar studies of fission barriers in actinides have been performed 
in Refs.\ \cite{LZZ.12,PNLV.12} using the RMF+BCS framework with 
the PC-PK1 and DD-PC1 parametrizations of the RMF Lagrangian. The 
accuracy of the description of the heights of inner fission barriers 
in these calculations is comparable with the one obtained in 
Ref.\ \cite{AAR.10}. The calculations of Refs.\ \cite{LZZ.12,PNLV.12} 
also include the results for outer fission barriers where the effects 
of octupole deformation (and triaxiality [Ref.\ \cite{LZZ.12}]) are 
taken into account. They agree reasonably well with experimental 
data.

  As compared with axially symmetric calculations, the inclusion 
of triaxiality has drastically improved the accuracy of the description 
of inner fission barriers in all model calculations (see Ref.\ 
\cite{AAR.12} and references quoted therein). Fig.\ 
\ref{N-dep}\footnote{Similar figure but as a function of 
proton number $Z$ is shown in Ref.\ \cite{AAR.12}.} shows 
that the state-of-the-art calculations within different theoretical
frameworks (including CDFT) are characterized by comparable 
accuracy (the $\delta$-values) of the description of inner fission 
barriers. Note that this comparison covers only the results of 
systematic triaxial calculations of even-even Th, U, Pu, Cm and 
Cf nuclei. Recent calculations within Skyrme DFT are also 
characterized by similar accuracy of the description of fission 
barriers \cite{SDFT.12,SDFT}.

  Minor differences between the approaches in the obtained 
average deviations per barrier (Fig.\ \ref{N-dep}) are 
not important considering the considerable uncertainties in the 
extraction of inner fission barrier heights from experimental data 
(see discussion in Ref.\ \cite{AAR.12-int}). However, the similarity 
of the average trends of these deviations (shown by thick dashed 
lines in Fig.\ \ref{N-dep}) as a function of neutron and proton 
numbers is more important considering the differences in underlying 
mean fields and in the treatment of pairing correlations. At present, 
no clear explanation for these trends is obtained 
\cite{AAR.12,AAR.12-int}. 

  The investigations of Refs.\ \cite{AAR.10,LZZ.12,PNLV.12} clearly 
indicate that good description of fission barriers in actinides in 
the CDFT framework is obtained with the parametrizations fitted with 
no information on fission barriers.  On the
contrary, successful description of fission barriers in 
actinides within non-relativistic DFT is based on the 
parametrizations which explicitly use either fission barrier 
heights (SkM* in Skyrme DFT and D1S in Gogny DFT) or fission
isomer excitation energies (UNEDF1 in Skyrme DFT); the latter
being strongly correlated with inner fission barrier height.

%%%%%%%%%%%%%%%%%%%%%%%%%%%%%%%%%%%%%%%%%%%%%%%%%%%%%%%%%%%%%%%%
\section{Fission barriers in superheavy nuclei.}
%%%%%%%%%%%%%%%%%%%%%%%%%%%%%%%%%%%%%%%%%%%%%%%%%%%%%%%%%%%%%%%

%%%%%%%%%%%%%%%%%%%%%%%%%%%%%%%%%%%%%%%%%%%%%%%%%%%%%%%%%%%%%%%%%%%%
\begin{figure}[h]
\centering
\includegraphics[width=8.0cm]{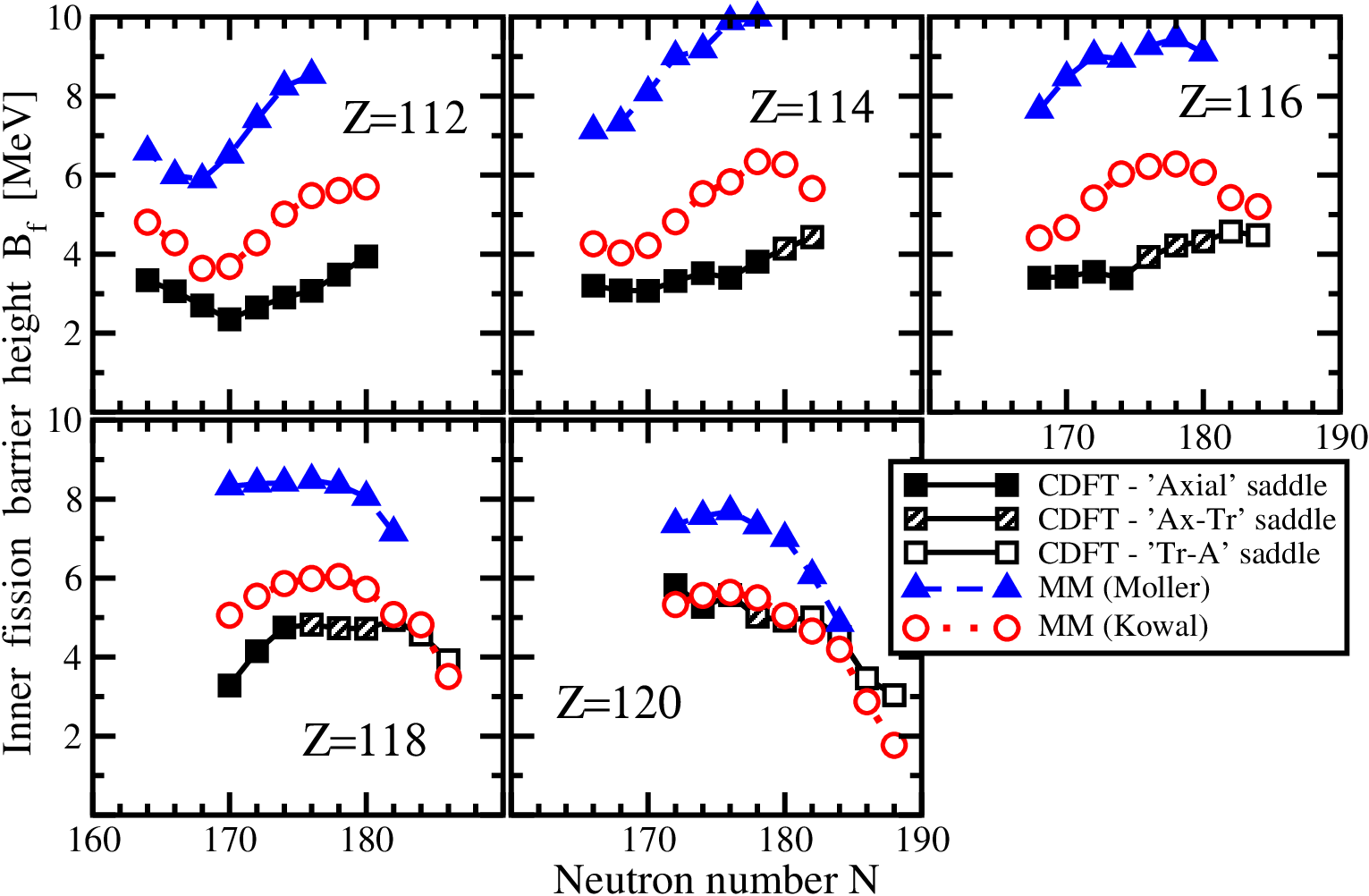}
\vspace{-0.0cm}
\caption{Inner fission barrier heights $B_f$ as a function
of neutron number $N$. The results of the MM calculations 
are taken from Ref.\ \protect\cite{MSI.09} (labeled as 'MM (M{\"o}ller)') 
and Ref.\ \protect\cite{KJS.10} (labeled as 'MM (Kowal)'). The position 
of inner fission barrier saddle in deformation space varies as 
a function of particle number. The labeling of Ref.\ \protect\cite{AAR.12} 
is used is order to indicate whether the saddle is axial (labeled as
'Axial'), has small  ($\gamma \sim 10^{\circ}$, labeled as 'Ax-Tr')
or large ($\gamma \sim 25^{\circ}$, labeled as 'Tr-A')
$\gamma-$deformations in the RMF+BCS calculations.}
\label{FB-SHE}
\end{figure}
%%%%%%%%%%%%%%%%%%%%%%%%%%%%%%%%%%%%%%%%%%%%%%%%%%%%%%%%%%%%%%%%%%%

  Fig.\ \ref{FB-SHE} shows how the models which have been 
bench-marked in a systematic way in the actinides (see Fig.\ 
\ref{N-dep}) extrapolate to the region of superheavy nuclei.
Note that the results of the MM calculations of Ref.\ 
\cite{KJS.10} labeled as 'MM(Kowal)' are not shown on 
Fig.\ \ref{N-dep}. However, they describe inner fission barriers 
of actinides very accurately. One can see that the model 
predictions vary wildly; the difference in inner fission barrier 
heights between different models reaches 6 MeV in some nuclei. 
This is despite the fact that these models describe the inner
fission barriers in actinides with comparable level of accuracy. 
The more surprising fact is that the prediction of two MM models 
differ so substantially; in reality the 'MM (Kowal)' model 
predictions are closer to the CDFT ones than to the 'MM (M{\"o}ller)' 
predictions.

%%%%%%%%%%%%%%%%%%%%%%%%%%%%%%%%%%%%%%%%%%%%%%%%%%%%%%%%%%%%%%%
\section{Some open questions}
%%%%%%%%%%%%%%%%%%%%%%%%%%%%%%%%%%%%%%%%%%%%%%%%%%%%%%%%%%%%%%%

 The results in the SHE region raise a number of questions which 
require further investigation. Among these, for example, are

\begin{itemize}

\item
  A. How the pairing evolves with deformation? How well the 
pairing strength is defined for SHE?

\item
  B. How the accuracy of the description of the energies of the 
single-particle states and its evolution with deformation affects 
the results of the calculations?
\end{itemize}

 %%%%%%%%%%%%%%%%%%%%%%%%%%%%%%%%%%%%%%%%%%%%%%%%%%%%%%%%%%%%%%%%%%%%
\begin{figure}[h]
\centering
\includegraphics[width=8.0cm]{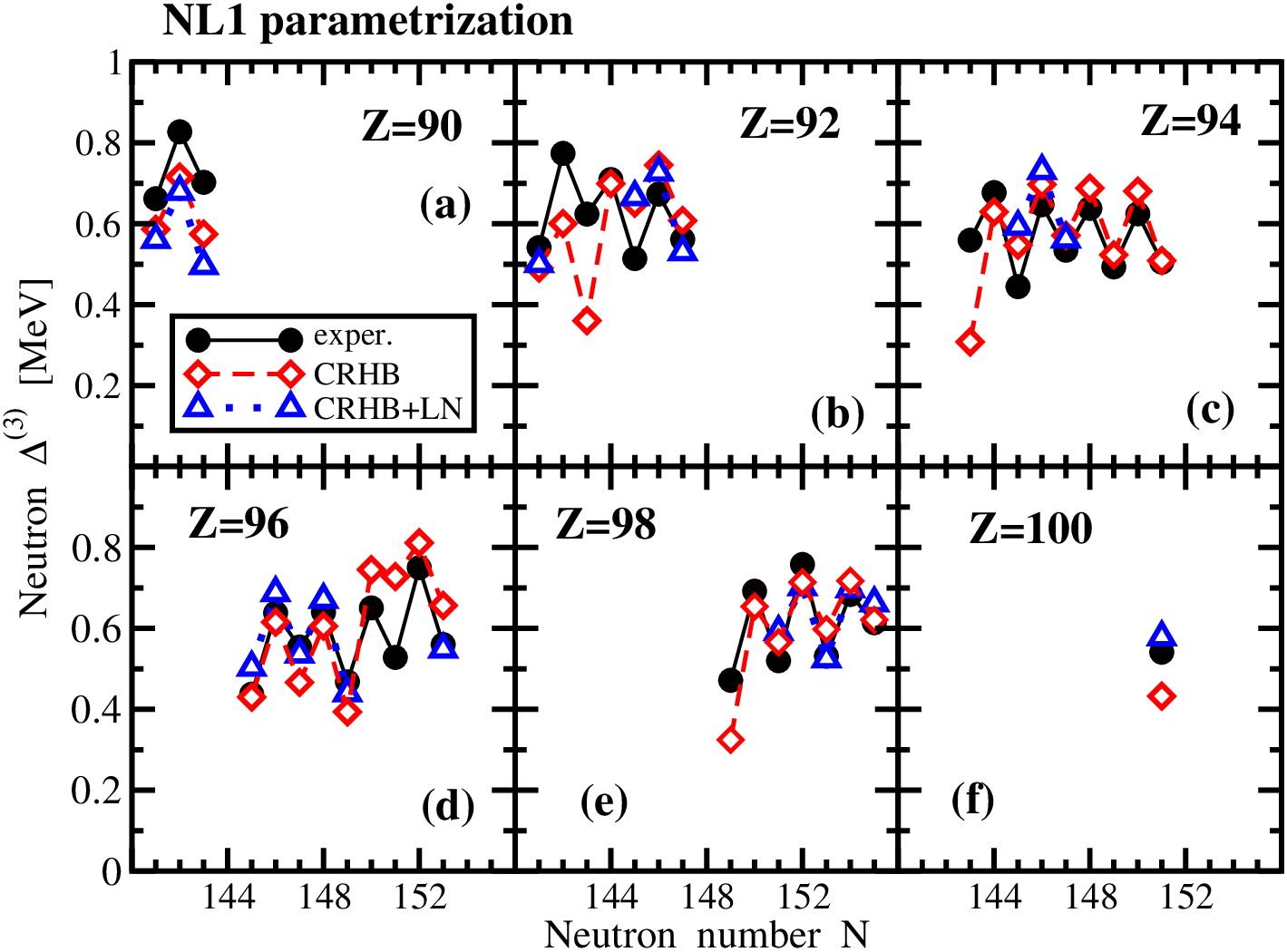}
\vspace{-0.1cm}
\caption{Experimental and calculated neutron three-point indicators 
$\Delta^{(3)}_{\nu}(N)$ as a function of neutron number $N$. The 
results of the CRHB and CRHB+LN calculations with the NL1 
parametrization are shown.}
\label{Delta3N-NL1}
\end{figure}
%%%%%%%%%%%%%%%%%%%%%%%%%%%%%%%%%%%%%%%%%%%%%%%%%%%%%%%%%%%%%%%%%%%

  I will concentrate on question A. Table IV in Ref.\ \cite{AAR.12} 
shows the large variety of pairing prescriptions used in the 
calculations of fission 
barriers during last decade. The question definitely emerges 
how well they describe the evolution of pairing with deformation
and extrapolate towards SHE. 
These features are important for a quantitative description of 
fission barrier heights which according to Ref.\ \cite{KALR.10} 
sensitively depend on  pairing properties. Although some attempts 
were made in 70ies to extract the information on pairing 
properties at fission saddles of actinides \cite{BH.74}, they 
did not lead to reliable estimates. It is even more difficult
to get information on pairing properties of SHE due to the lack
of relevant experimental data. As a consequence, as accurate as 
possible description of pairing properties in actinides and light 
SHE should be considered as a necessary condition for further 
extrapolation to heavier SHE.

%%%%%%%%%%%%%%%%%%%%%%%%%%%%%%%%%%%%%%%%%%%%%%%%%%%%%%%%%%%
\begin{figure}[ht]
\centering
\includegraphics[width=8.0cm]{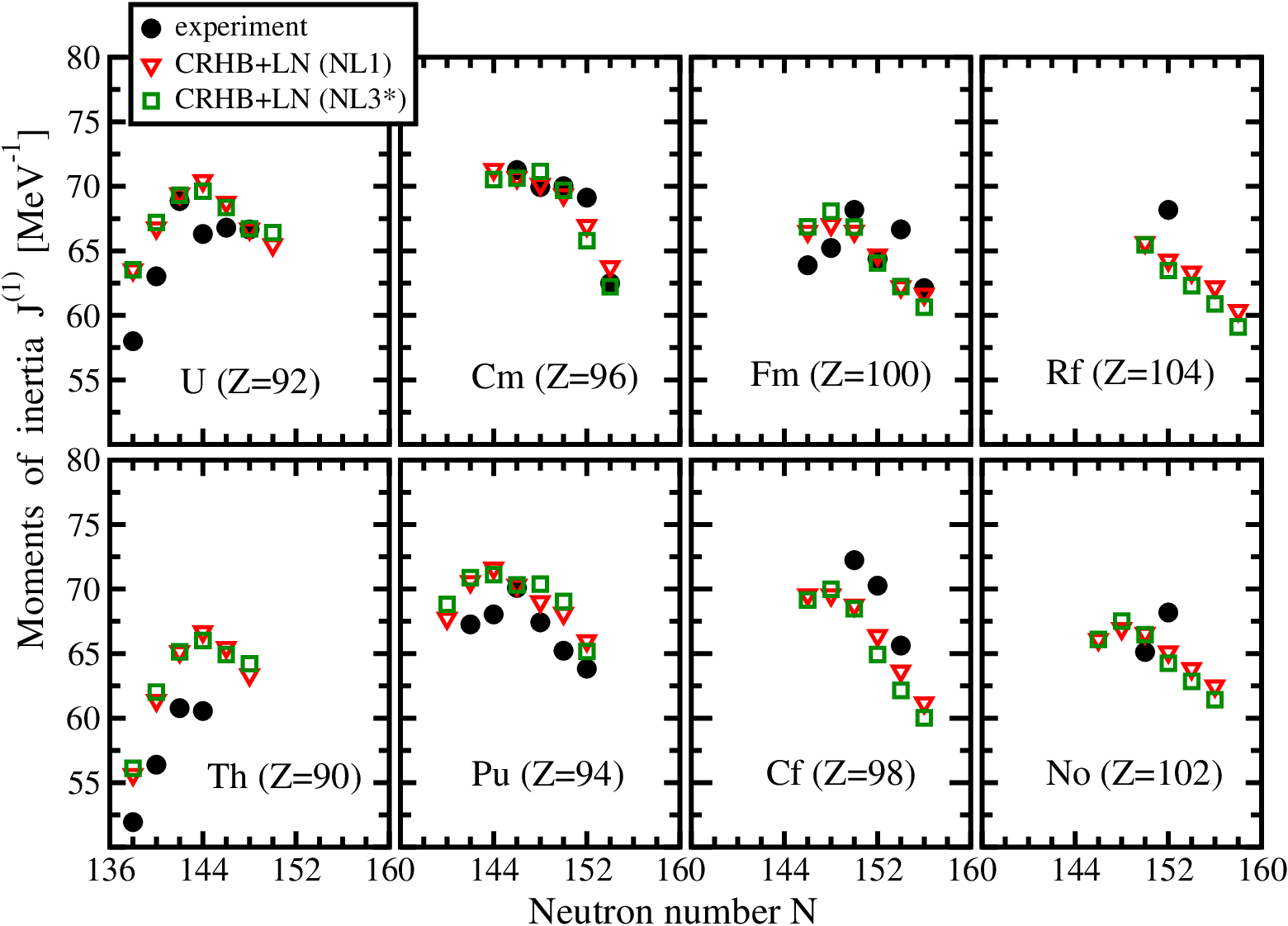}
\vspace{-0.2cm}
\caption{Calculated and experimental moments of inertia at 
low spin. }
\label{J1-low}
\end{figure}
%%%%%%%%%%%%%%%%%%%%%%%%%%%%%%%%%%%%%%%%%%%%%%%%%%%%%%%%%%%

  In order to address these questions, a comprehensive
study of pairing and rotational properties of actinides is performed 
in Ref.\ \cite{AA.13} in the cranked relativistic Hartree-Bogoliubov
(CRHB) framework. In this approach, the 
Brink-Booker part of phenomenological non-relativistic Gogny-type 
finite range interaction 
\begin{eqnarray}
V^{pp}(1,2) =  f \sum_{i=1,2} e^{-[({\bm r}_1-{\bm r} _2)/\mu_i]^2}  
(W_i+B_i P^{\sigma}- H_i P^{\tau} - M_i P^{\sigma} P^{\tau})
\label{Vpp}
\end{eqnarray}
is used in pairing channel. In addition, an approximate particle number 
projection by means of Lipkin-Nogami (LN) method is employed.
The analysis indicates that the pairing strength has to be decreased as
compared with original D1S Gogny force; the scaling factor $f$ is equal 
to 0.899 and 0.9147 for the NL3* and NL1 parametrizations, respectively. 
Only with these scaling factors, the moments of inertia and odd-even 
mass staggerings (the $\Delta^{(3)}$ indicators) can be well described
in the normal-deformed (ND) minimum (Figs.\ 
\ref{Delta3N-NL1} and \ref{J1-low}).

%%%%%%%%%%%%%%%%%%%%%%%%%%%%%%%%%%%%%%%%%%%%%%%%%%%%%%%%%%%%%%%%%%%%%%%%
\begin{table}
\tbl{Experimental and theoretical charge quadrupole moments 
$Q$  of SD fission isomers. The results of the CRHB+LN calculations 
with the NL1 and NL3* parametrizations are presented. Experimental 
data for the U and Pu isotopes are taken from Ref.\ \protect\cite{H.89},
while the one for $^{242}$Am from Ref.\ \protect\cite{Am242}.}
{\begin{tabular}{ccccccc} \hline
                   & $^{236}$U & $^{238}$U & $^{236}$Pu & $^{239}$Pu &
                   $^{240}$Pu & $^{242}$Am \\ \hline
$Q^{exp}$ ($e$b) & $32\pm 5$ & $29\pm 3$ & $37\pm 10$ & $36\pm 4$ &
                   & $35.5\pm 1.0_{st}\pm 1.2_{mod}$ \\
$Q^{\rm NL1}$ ($e$b)  &  35.8 & 37.3 & 36.1 & & 38.2 & \\
$Q^{\rm NL3*}$ ($e$b) &  33.9 & 33.7 & 34.8 & & 34.9 & \\ \hline 
\end{tabular}}
\label{QT-SD}
\end{table}
%%%%%%%%%%%%%%%%%%%%%%%%%%%%%%%%%%%%%%%%%%%%%%%%%%%%%%%%%%%%%%%%%%%%%%

  Fission isomers provide only available tool to estimate the evolution 
of pairing with deformation in actinides. Such an estimate is available 
only through the study of rotational properties of $^{236,238}$U and 
$^{240}$Pu nuclei; these are only nuclei for which superdeformed (SD) 
rotational bands were experimentally measured. The calculated charge 
quadrupole moments $Q$ are compared with available experimental data 
in Table \ref{QT-SD}. The CRHB+LN(NL3*) results of the calculations come 
reasonably close to experiment.  The CRHB+LN(NL1) results are also 
not far away from experimental data but they substantially overestimate 
experimental $Q$ value in $^{238}$U.

\begin{figure}[h]
\centering
\includegraphics[width=8.0cm,angle=0]{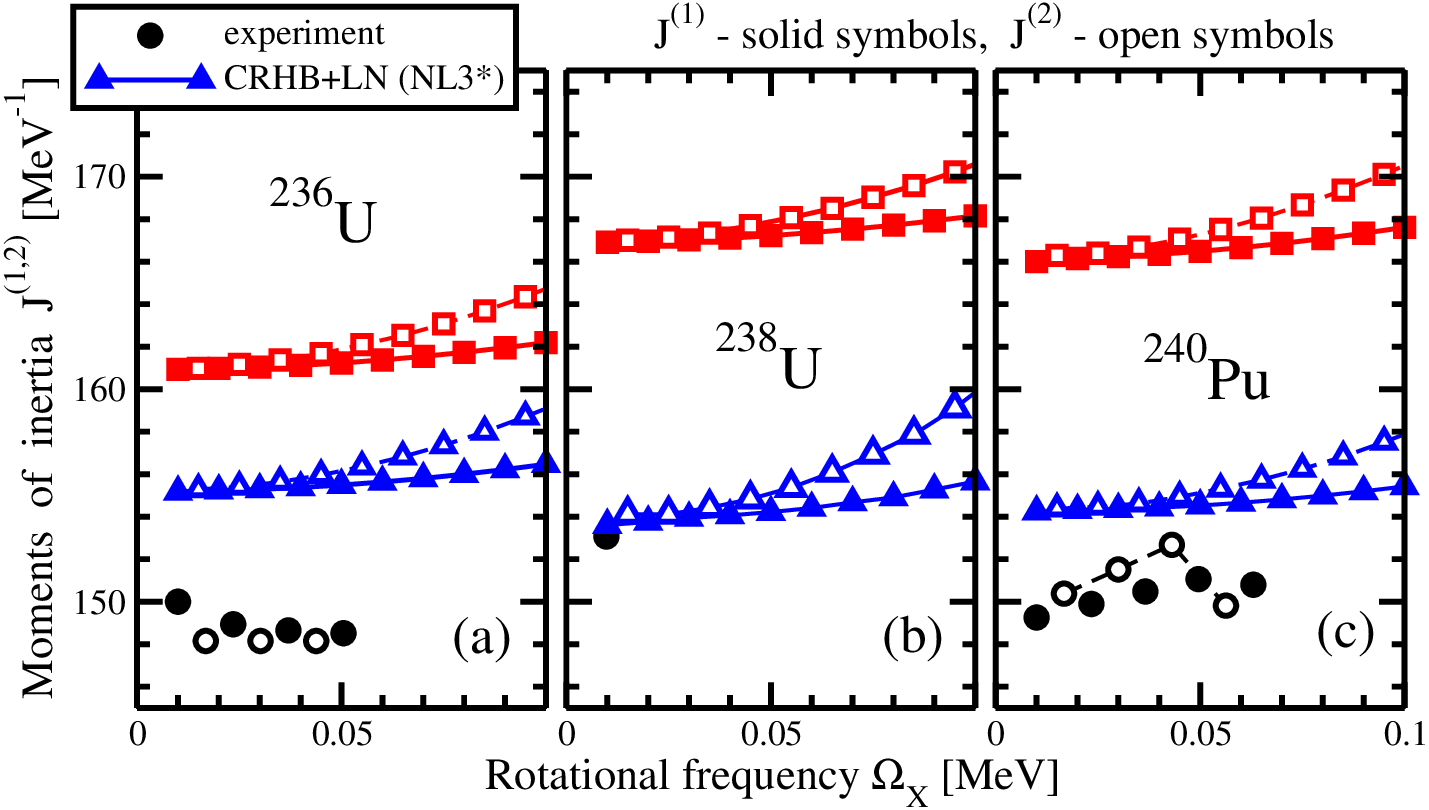}
\vspace{-0.2cm}
\caption{Experimental and calculated kinematic $(J^{(1)})$ and 
dynamic $(J^{(2)})$ moments of inertia of SD rotational bands 
in $^{236,238}$U and $^{240}$Pu. The notation of the lines and 
symbols is given in the figure.} 
\label{J1-SD}
\end{figure}

  The experimental kinematic moments of inertia are best described by 
the NL3* parametrization; the deviation from experiment does not exceed 
3.4\% (Fig.\ \ref{J1-SD}). The fact that the moments of inertia of 
rotational structures in two different minima (ND and SD) are accurately 
described with the same pairing strength strongly suggests that the 
evolution of pairing correlations with deformation is properly described 
in the CRHB+LN(NL3*) framework by the Gogny D1S pairing.

%%%%%%%%%%%%%%%%%%%%%%%%%%%%%%%%%%%%%%%%%%%%%%%%%%%%%%%%%%%%%%%%%%
%%%%%%%%%%%%%%%%%%%%%%%%%%%%%%%%%%%%%%%%%%%%%%%%%%%%%%%%%%%%%%%%%%
\vspace{-0.3cm}

  However, this is not always the case since the CRHB+LN(NL1) calculations
substantially overestimate the experimental moments of inertia in the 
SD minimum [for example, by 11.3\% in $^{240}$Pu] (Fig.\ \ref{J1-SD})
while they reproduce the low-spin moments of inertia in the ND minimum 
with the same level of accuracy as the CRHB+LN(NL3*) calculations (see 
Fig.\ \ref{J1-low}). It turns out that reasonable description of the 
moments of inertia at SD can be achieved in the CRHB+LN(NL1) calculations 
only if scaling factor $f\approx 1.0$ is used. This clearly indicates 
that even the pairing force carefully fitted to experimental data at 
ND does not guarantee accurate description of pairing at SD (and as a 
consequence also at fission  saddle).

  The origin of such behavior is not completely clear but the 
difference between the CRHB+LN(NL3*) and CRHB+LN(NL1*) results for 
$J^{(1)}$ at SD may also partially originate from the differences in 
the single-particle structures at superdeformation  obtained with the 
NL3* and NL1 parametrizations. The $Q$ values obtained in 
the CRHB+LN(NL1) calculations are always higher than the ones 
for CRHB+LN(NL3*), which also may be a reason why the 
CRHB+LN(NL1) calculations systematically overestimate kinematic 
moments of inertia at SD.

% Although fission isomers in 
% actinides have been observed more than 50 years ago their rotational 
% and single-particle properties 
% are significantly less known experimentally than in other regions of 
% superdeformation. For example, no reliable experimental data on the 
% single-particle states in odd-mass actinides exist.

%%%%%%%%%%%%%%%%%%%%%%%%%%%
\section{Conclusions}
%%%%%%%%%%%%%%%%%%%%%%%%%%%

  The review of the recent applications of covariant density functional
theory to the description of fission barriers in actinides and superheavy
nuclei is presented. Inner fission barriers of actinides are described 
with comparable accuracy in the MM and DFT (including CDFT) calculations.
However, the predictions for fission barriers of SHE within different 
frameworks vary drastically indicating the need for better parametrizations 
of different channels of the models. In particular, a better pairing force 
and its better parametrization is needed. An effort to improve the Gogny 
pairing force in the RHB framework, the first results of which are 
presented here, is currently in progress. 

%%%%%%%%%%%%%%%%%%%%%%%%%%%%%%%
\section*{Acknowledgments}
%%%%%%%%%%%%%%%%%%%%%%%%%%%%%%%

  This work has been supported by the U.S. Department of Energy 
under the grant DE-FG02-07ER41459. I would like to express my gratitude 
to P.\ Ring, H.\ Abusara and O.\ Abdurazakov for their contributions
into this project.

\end{document}